\documentclass[a4paper,11pt]{article}
\usepackage{pos}

\title{Damping of density oscillations from bulk viscosity in quark matter}

\author{Jos\'e Luis Hern\'andez$^{1,2,3}$, Cristina Manuel$^{1,2}$ and Laura Tolos$^{1,2,4}$  }
\affiliation{
$^1$Institute of Space Sciences (ICE, CSIC), Campus UAB,  Carrer de Can Magrans, 08193 Barcelona, Spain\\
$^2$Institut d'Estudis Espacials de Catalunya (IEEC), 08034 Barcelona, Spain \\
$^3$Facultat de Física, Universitat de Barcelona, Martí i Franquès 1, 08028 Barcelona, Spain.\\
$^4$ Frankfurt Institute for Advanced Studies, Ruth-Moufang-Strasse 1, 60438 Frankfurt am Main, Germany
}

\emailAdd{hernandez@ice.csic.es}
\emailAdd{cmanuel@ice.csic.es}
\emailAdd{tolos@ice.csic.es}

\abstract{In this contribution, we extend the discussion about the calculation of the  bulk viscosity of quark matter in the normal phase due to electroweak processes and its effect on the damping of baryon density oscillations that might occur in the coalescence of two compact stars.  Employing the EoSs from the MIT bag model and perturbative quantum chromodynamics (pQCD) up to $\mathcal{O}(\alpha_s)$, we analyze our results varying densities in the range of temperatures from 0 to 10 MeV for frequencies around 1 kHz. Our estimates show that bulk viscous effects might play a relevant role during the postmerger stage if the system reaches a deconfined quark matter phase.}

\FullConference{10th International Conference on Quarks and Nuclear Physics (QNP2024)\\
8-12 July, 2024\\
Barcelona, Spain\\}

\begin{document}
\maketitle

\section*{Introduction}
From the theoretical point of view, a lot of effort has been invested in elucidating the composition of compact stars. Conventional approaches involve the study of different microscopic models and their associated equations of state (EoSs), the dynamical processes that occur in their interior, and their response to external perturbations. In the latter case, one resorts to transport theory to obtain the rate at which conserved quantities such as energy, momentum, mass, charge and others are transferred from one region to another within the medium by computing the transport coefficients (viscosities, relaxation times, conductivities and others)~\cite{Schmitt_2018}.

The value of the transport coefficients is determined by the microscopic composition and the dominant interactions of the constituents. Concerning the bulk viscosity, there are indications that bulk viscous dissipation may be physically important in the survival time of the post-merger object a few milliseconds after the coalescence of two neutron stars~\cite{Alford_2018}. Thus, the inclusion of bulk viscosity on numerical simulations of mergers might be important in its postmerger dynamics.

\section*{Damping of density oscillations in three-flavour quark matter}
In this section, we revisit the calculation of the bulk viscosity generated in unpaired three-flavour quark matter (in the neutrino-transparent regime) based on Ref.~\cite{PhysRevD.109.123022}. We study its implications on the damping of baryonic density oscillations and analyze our results using two EoSs.
 
Due to the expansion or compression of the medium, the system is subject to instantaneous departures of chemical equilibrium which can be studied in terms of chemical imbalances: $\mu_1 =\mu_s -\mu_d$, $\mu_2 =\mu_s -\mu_e -\mu_u,$ and $ \mu_3 =\mu_d -\mu_e -\mu_u$, where $\mu_1=\mu_2=\mu_3=0$ in beta equilibrium at low temperatures.

For unpaired three-flavour quark matter at low temperatures where the medium is transparent to neutrinos, the nonleptonic electroweak processes ($u+d \leftrightarrow u+s$) and the semileptonic processes ($u+e^- \rightarrow d+\nu_e, \, d \rightarrow u+e^-+\bar{\nu}_e, \, u+e^- \rightarrow s+\nu_e , \,
s \rightarrow u+e^-+\bar{\nu}_e$) might re-establish the chemical equilibrium during the timescale of density oscillations in the postmerger stage.

Out of equilibrium these processes generate source and sink terms, which break the conservation laws of the individual particle currents. Imposing the local charge neutrality and using the baryon number density $n_B$, the evolution of the particle number densities can be studied using a set of equations for the divergence of the particle density current of strange quarks ($n_s u^\mu$) and electrons ($n_e u^\mu$) at linear order in the chemical imbalances
\begin{equation}
    \partial_\mu(u^\mu n_s)=-\lambda_1 \mu_1 -\lambda_2 \mu_2, \quad \partial_\mu(u^\mu n_e)=\lambda_2 \mu_2 +\lambda_3 \mu_3, \label{eq.evonsandne}
\end{equation}
where $\lambda_1, \, \lambda_2, \, \text{and} \, \lambda_3$ are the coefficients of the following combinations of rates expanded at linear order in the chemical imbalances:
\begin{eqnarray}
     &&\Gamma_{s+u \rightarrow d+u} - \Gamma_{d+u \rightarrow s+u}=\mu_1 \lambda_1, \label{rate1} \\
     &&\Gamma_{s \rightarrow u+e+\bar{\nu}_e} - \Gamma_{u+e \rightarrow s+\nu_e}=\mu_2 \lambda_2,  \label{rate2}\\
     && \Gamma_{d \rightarrow u+e+\bar{\nu}_e} - \Gamma_{u+e \rightarrow d+\nu_e}=\mu_3 \lambda_3.  \label{rate3}
\end{eqnarray}

These rates are relevant in the estimation of the bulk viscosity and have been calculated at tree level in different limits for the chemical potentials and masses of the involved particles (see references in Ref~\cite{PhysRevD.109.123022}). In this work, for the nonleptonic channel, we use the low temperature limit assuming massless quarks, $\mu_u =\mu_d$ and  we assume that the mismatch between the strange and down quark chemical potentials is not too large. For the semileptonic channels we consider that the temperature, the quark chemical imbalances and the strange quark mass are small compared with the quark chemical potentials. 

Deviations of the particle number density of their equilibrium values can be computed from Eq.\eqref{eq.evonsandne} assuming that the oscillating parts of the particle density are proportional to $e^{i\omega t}$, being $\omega$ the angular frequency of the oscillation. 

At first-order hydrodynamics, the bulk viscosity is given by
\begin{equation}
    \zeta \equiv -\frac{\text{Re}[\Pi]}{\theta},
\label{bulkdef}
\end{equation}
where $\Pi$ is the bulk scalar, which quantifies the modification of the pressure from its value in chemical equilibrium induced by the deviations of the particle number densities and chemical potentials out of equilibrium
\begin{equation}
    \Pi=(C_1 -C_2)\delta n_e' +C_1 \delta n_s'.
\end{equation}
Here we define  
\begin{equation}
C_1 \equiv n_{s,0}A_s-n_{d,0}A_d, \quad C_2 \equiv n_{s,0}A_s-n_{u,0}A_u-n_{e,0}A_e ,
\end{equation}
where $n_{j,0}$ are the instantaneous values of the particle number densities in chemical equilibrium and $A_j$ are the susceptibilities of the constituent particles written in terms of the partial derivatives of the particle's chemical potentials with respect to the particle number density $A_{ij}=\left(\partial \mu_i/\partial n_j \right)_{n_{k\neq j},T}$.

As a result, the bulk viscosity due to electroweak processes in the neutrino-transparent regime for three-flavour quark matter in the normal phase is given by
\begin{equation}
    \zeta = \frac{\kappa_1 +\kappa_2 \omega^2}{\kappa_3 +\kappa_4 \omega^2 +\omega^4},
    \label{finalbulkviscosity}
\end{equation}
with the following definitions: $A\equiv  A_e + A_u$, $\lambda_Q\equiv \lambda_1 \lambda_2+ \lambda_1 \lambda_3+ \lambda_2 \lambda_3 $,
\begin{eqnarray}
    \kappa_1 &\equiv& \lambda_Q \{ C_1^2[(A+A_d)[A(\lambda_2 +\lambda_3)+A_d \lambda_3]-A_d (A\lambda_2 -A_d \lambda_1)] \nonumber \\
    &-&2C_1(C_1 -C_2) [A_d[(A_d +A_s)\lambda_1 +(A+A_d)\lambda_3]-AA_s\lambda_2] \nonumber \\
    &+&(C_1-C_2)^2[\lambda_1 (A_d+A_s)^2+\lambda_2A_s^2+\lambda_3 A_d^2] \}, \\
    \kappa_2 &\equiv&  \lambda_1 C_1^2+\lambda_2 C_2^2 +\lambda_3(C_1 -C_2)^2 ,\\    
    \kappa_3 &\equiv& \lambda_Q^2 [A(A_s +A_d)+A_d A_s]^2, \\
    \kappa_4 &\equiv& [(A_d +A_s)\lambda_1 +A_s \lambda_2]^2 +2(A\lambda_2 -A_d \lambda_1)[A_s (\lambda_2 +\lambda_3)-(A_d +A_s)\lambda_3] \nonumber \\
    &+&[A_d \lambda_3 +A(\lambda_2 +\lambda_3)]^2.
\end{eqnarray}

The damping time of baryon number density oscillations in a bulk viscous medium can be estimated assuming small density oscillations described by $\delta n_B= \Delta n_{B} e^{i \omega t}$, being $\Delta n_{B}$ the amplitude of the oscillation.
The damping time is defined by $\tau_\zeta \equiv \epsilon/(d\epsilon/dt)$ where $\epsilon$ is the energy density stored in a baryonic density oscillation and $d\epsilon/dt$ is the energy dissipation time. It can be computed in terms of the energy density of the system $\varepsilon$ as follows~\cite{PhysRevD.39.3804,Alford_2018}
\begin{equation}
    \tau_\zeta = \frac{n_{B,0}^2}{\omega^2 \zeta} \left(\frac{\partial^2 \varepsilon}{\partial n_B^2} \right)_{n_s,n_e,s}.
    \label{dampingtime}
\end{equation}

\begin{figure}
\includegraphics[width=0.75\textwidth]{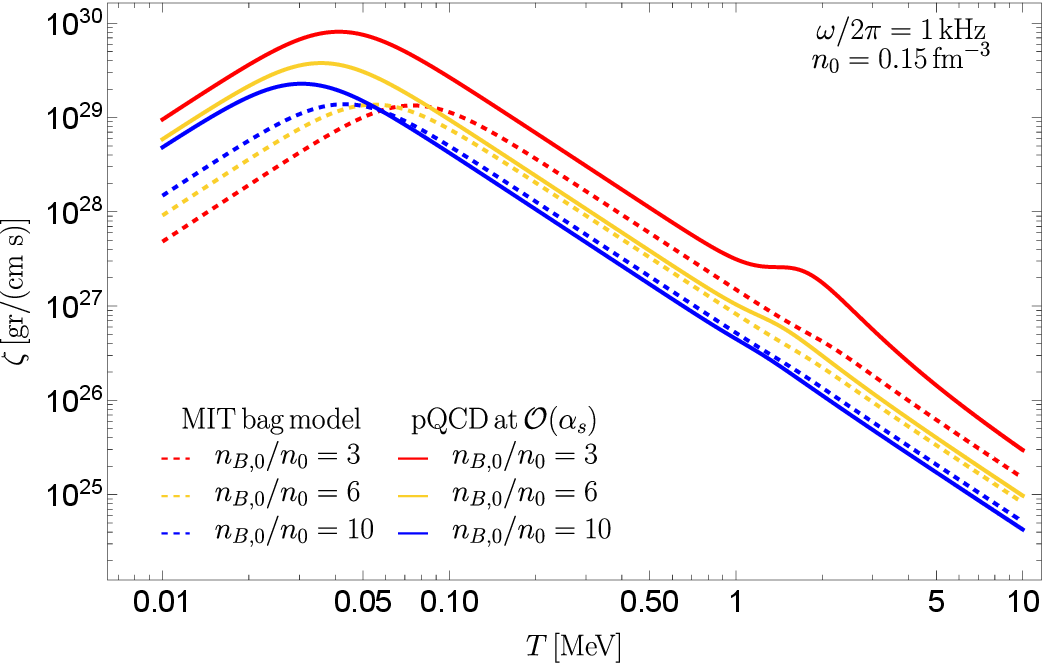}
\centering
\caption{Bulk viscosity of three-flavor quark matter in the neutrino-free regime for different normalized baryon number densities at $\omega/2\pi=1$ kHz using the MIT bag model (dashed lines) at $m_s=150$ MeV and pQCD up to $\mathcal{O}(\alpha_s)$ with running $m_s$ and $\alpha_s$ (continuous lines).}
\label{fig:bv}
\end{figure}

\begin{figure}
\includegraphics[width=0.75\textwidth]{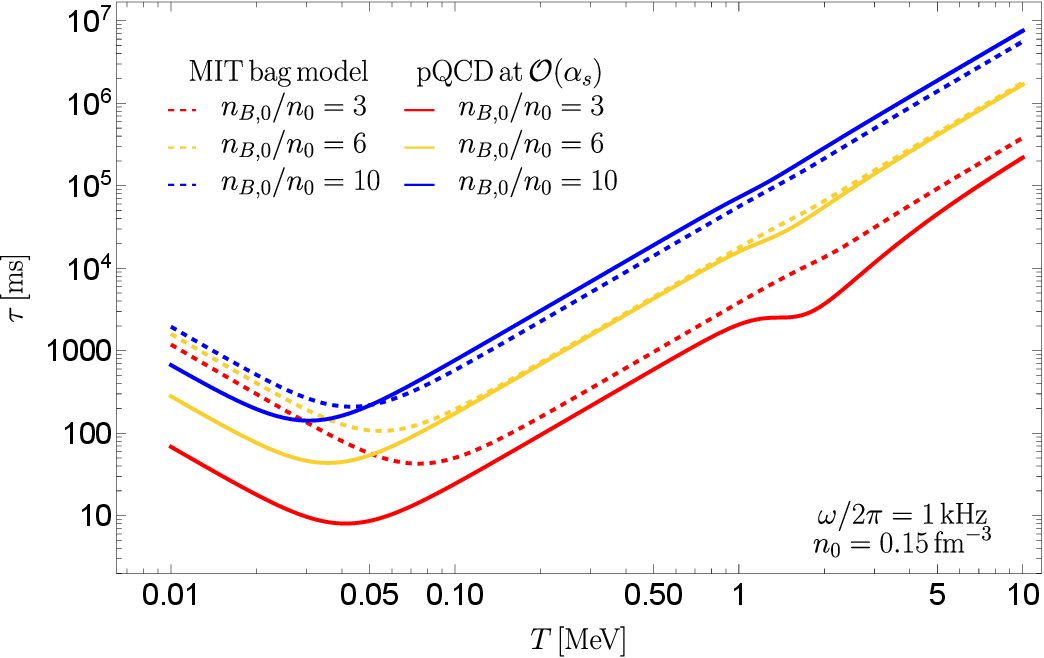}
\centering
\caption{Damping times of baryon density oscillations in unpaired quark matter with similar input parameters as for the bulk viscosity in Fig.~\ref{fig:bv}.}
\label{fig:tau}
\end{figure}

In Figs.~\ref{fig:bv} and \ref{fig:tau} we show some numerical results using the EoSs of the MIT bag model and pQCD in the zero temperature limit varying the density as multiples of the nuclear saturation density $n_0=0.15$ $\text{fm}^{-3}$. For both models we assume that the light quarks are massless and consider only the contribution of the strange quark mass. It is worth to note that the bulk viscosity and the damping time depend strongly on the value of the strange quark mass, more details about this can be found in Ref.~\cite{PhysRevD.109.123022}. In Fig.~\ref{fig:bv} we find that the maximum of the bulk viscosity is in the range of temperature from $0.01$ to $0.1$ MeV at $\omega/2\pi=1$ kHz, while the associated minimal damping times of the density oscillations at those temperatures lie in the range of few to hundreds milliseconds, as can be seen in Fig.~\ref{fig:tau}. This timescale is similar to the baryonic density oscillations' timescale determined by numerical simulations of neutron star mergers (analogous conclusions have been obtained for dense nuclear matter at higher temperatures, see~\cite{Harris_2024} and its list of references), thereby opening up the possibility of bulk viscosity being important during the postmerger if the medium is composed of normal quark matter. 
\subsection*{Conclusions}

We adressed the bulk viscosity and the damping time of density oscillations in normal quark matter using the EoSs of the MIT bag model and pQCD up to $\mathcal{O}(\alpha_s)$, and explored their dependence on the baryon number density, temperature and strange quark mass, being the latter the most important input parameter of these quantities. 

We find that the maximum value of the bulk viscosity, which produce the shortest damping times of the density oscillations (in the order of the few to several hundreds of milliseconds, depending on values of the density and the strange quark mass), occurs at temperatures in the range from 0.01  to 0.1  MeV. The precise value depends on the  EoS describing  quark matter and the rates of the electroweak processes involved.  Our results suggest that baryonic density oscillations might be damped if the medium contains deconfined quark matter, making bulk viscosity relevant in the postmerger phase after the collision of two neutron stars.

\subsection*{Acknowledgments}
We acknowledge support from the program Unidad de Excelencia María de Maeztu CEX2020-001058-M, from Projects No. PID2019-110165GB-I00 and No. PID2022-139427NB-I00 financed by the Spanish MCIN/AEI/10.13039/501100011033/FEDER, UE (FSE+),  as well as from the Generalitat de Catalunya under Contract No. 2021 SGR 171,  by  the EU STRONG-2020 project, under the program  H2020-INFRAIA-2018-1 Grant Agreement No. 824093, and by the CRC-TR 211 \lq \lq Strong-interaction matter under extreme conditions" Project No. 315477589-TRR 211.


\begin{thebibliography}{99}

\bibitem{Schmitt_2018}
Schmitt, Andreas and Shternin, Peter, Reaction Rates and Transport in Neutron Stars, \textit{The Physics and Astrophysics of Neutron Stars}, Springer International Publishing, 2018, 455–574.

\bibitem{Alford_2018}
Alford, Mark G. and Bovard, Luke and Hanauske, Matthias and Rezzolla, Luciano and Schwenzer, Kai, Viscous Dissipation and Heat Conduction in Binary Neutron-Star Mergers, \textit{Phys. Rev. Lett.}, 120 (2018), 041101.

\bibitem{PhysRevD.109.123022}
Hern\'andez, Jos\'e Luis and Manuel, Cristina and Tolos, Laura, Damping of density oscillations from bulk viscosity in quark matter,  \textit{Phys. Rev. D}, 109 (2024), 123022.

\bibitem{PhysRevD.39.3804}
Sawyer, Raymond F., Bulk viscosity of hot neutron-star matter and the maximum rotation rates of neutron stars,  \textit{Phys. Rev. D}, 39 (1989), 3804--3806.

\bibitem{Harris_2024}
Harris, Steven P., Bulk Viscosity in Dense Nuclear Matter, \textit{Nuclear Theory in the Age of Multimessenger Astronomy}, CRC Press, 2024, 215–260.

\end{thebibliography}
\end{document}